\documentstyle[12pt]{article}
\begin{document}
\def\a{\alpha}
\def\b{\beta}
\def\ch{\chi}
\def\d{\delta}
\def\e{\epsilon}
\def\f{\phi}
\def\g{\gamma}
\def\h{\eta}
\def\i{\iota}
\def\j{\psi}
\def\k{\kappa}
\def\l{\lambda}
\def\m{\mu}
\def\n{\nu}
\def\o{\omega}
\def\p{\pi}
\def\q{\theta}
\def\r{\rho}
\def\s{\sigma}
\def\t{\tau}
\def\u{\upsilon}
\def\x{\xi}
\def\z{\zeta}
\def\D{\Delta}
\def\F{\Phi}
\def\G{\Gamma}
\def\J{\Psi}
\def\L{\Lambda}
\def\O{\Omega}
\def\P{\Pi}
\def\S{\Sigma}
\def\U{\Upsilon}
\def\X{\Xi}
\def\T{\Theta}
\def\Ab{\bar{A}}
\def\gi{g^{-1}}
\def\li{{ 1 \over \l } }
\def\lb{\l^{*}}
\def\zb{\bar{z}}
\def\ub{u^{*}}
\def\Tb{\bar{T}}
\def\pp {\partial }
\def\pb {\bar{\partial }}
\def\be{\begin{equation}}
\def\ee{\end{equation}}
\def\ben{\begin{eqnarray}}
\def\een{\end{eqnarray}}
\begin{flushright} \ December \ 1996\\
SNUTP/96-123 \\
\end{flushright}
\hspace{.5cm}
\begin{center}
{\bf  The Field Theory of Generalized Ferromagnet on \\
the Hermitian Symmetric Spaces
\footnote{Talk given at 15th Symposium on Theoretical Physics:
"Field Theoretical Methods in Fundamental Physics",
Seoul, Korea, August 1996.}
 }\\
\end{center}

\begin{center}
Phillial Oh \\
Department of Physics, 
Sung Kyun Kwan Unversity\\
Suwon 440-746, 
Republic of Korea \\
E-mail:ploh@newton.skku.ac.kr\\
\end{center}
\vspace{1.0cm}
\begin{abstract}
We discuss the recent developments in the generalized 
continuous Heisenberg ferromagnet model formulated as a nonrelativistic 
field theory defined on the target space of the coadjoint orbits. 
Hermitian symmetric spaces are special because they provide completely  
integrable field theories in 1+1 dimension and self-dual Chern-Simons
solitons and vortices in 2+1 dimension.
\end{abstract}
\noindent 

\vspace{1.0cm}
Recently, an action principle of a nonrelativistic nonlinear 
sigma model with the target space of coadjoint orbits and
its coupling with the Chern-Simons gauge field was proposed 
\cite{park,park1}. The coadjoint orbits are 
naturally equipped with symplectic structure \cite{kiri} 
and this can be used to construct  the action for nonrelativistic 
field theories of generalized spins which are defined on them 
with arbitrary groups. The resulting models describe generalized 
Heisenberg ferromagnet in which the equation of motion 
satisfies the generalized  Landau-Lifshitz (LL) equation. 
The Hermitian symmetric spaces \cite{ford}
which are special types of the coadjoint orbits are especially interesting
because they provide completely integrable field theories in 1+1 
dimension \cite{park} and self-dual Chern-Simons solitons and vortices in 2+1 
dimension \cite{park1}. 
In this talk, I will present a review on the subject and discuss the related issues.
This work was done in collaboration with Q-Han Park.

We start with a brief summary of the phase space of the coadjoint orbits
and the Hermitian symmetric space.
Consider a cotangent bundle $T^*G\cong G\times {\cal G}^*$ \cite{arno}
of an arbitrary group $G$ which can be regarded as the phase space for
the  generalized spin  degrees of freedom. ${\cal G}^*$
is the dual of the Lie algebra ${\cal G}$ of the
group G. There is a natural canonical one form
\begin{equation}
\theta=2\mbox{Tr}~(xg^{-1}dg),
\label{theta}
\end{equation}
and symplectic structure
\begin{equation}
\omega=2\mbox{Tr}~ (x dg^{-1}\wedge dg),
\label{omega}
\end{equation}
where $x\in {\cal G}^*$ is assumed to be constant. Then, a 
natural symplectic left group action on $T^*G$ can be defined by
\cite{ohnucl}
\begin{equation}
G\times (G\times {\cal G}^*)\longrightarrow
G\times {\cal G}^*
\end{equation}
\begin{equation}(g^\prime, (g,a))\mapsto (g^\prime g,a).
\end{equation}
Let us consider the associated moment map
$ \rho: T^*G\rightarrow {\cal G}^* $ via
\begin{equation}
<X, \rho(m)>=m\left(\frac{d}{dt}\Big\vert_{t=0}
\exp tX\circ g\right),
\label{map}
\end{equation}
where $X\in {\cal G}$ and $m\in T^*_gG$ is a linear map of ${\cal G}\rightarrow {\bf R}$.
Then,  $\rho^A(m)\equiv <T^A, \rho(m)>$'s, where
$T^A$'s are the generator of ${\cal G}$, $[T^A, T^B]=f^{AB}_{\ \ C}
T^C$ with  $\mbox{Tr}(T^AT^B)=-1/2\eta_{AB}$,
realize the Lie algebra \cite{arno}:
\begin{equation}
\{\rho^A, \rho^B\}=f^{AB}_{\ \ C}\rho^C.\label{poisson}
\end{equation}

It is well known that $T^*G$ can be reduced with respect to
the above momentum map by the symplectic reduction and the reduced 
phase space is naturally  identifiable with the coadjoint orbit
${\cal O}_x\equiv G\cdot x\subset {\cal G}^*$:
\begin{equation}
\rho^{-1}(x)/G_x\cong G/G_x\cong  G\cdot x,
\end{equation}
where $G_x$ is the stabilizer group of the point $x$.
The symplectic structure (\ref{theta}), (\ref{omega}) naturally 
descend on the coadjoint orbit $G/G_x\equiv G/H$ . 
Also the reduced moment map (\ref{map}) becomes the   
generalized spin  degrees of freedom 
which can equivalently be expressed by
\begin{equation}
Q=\mbox{Ad}^*(g)x=gxg^{-1} \quad g \in G.
\end{equation}
Then, on the coadjoint orbit, the Eq. (\ref{map}) becomes
the generalizes  spin algebra \cite{arno,bak};
\begin{equation}
\{Q^A, Q^B\}=f^{AB}_{\ \ C}Q^C.
\label{spin}
\end{equation}
An explicit example of the above procedure with an arbitrary coadjoint
orbit of $G=SU(N)$ can be found in Ref. \cite{ohnucl}. 

Now, we  give a brief description  of Hermitian symmetric spaces 
\cite{ford} which are special type of coadjoint orbits.
 A symmetric space is a coset space $G/H$ for Lie groups 
whose associated Lie algebras ${\cal G}$ and $ {\cal H}$, 
with the decomposition
${\cal G} = {\cal H} \oplus {\cal M}$, satisfy the commutation relations,
\begin{equation}
[{\cal H} , ~ {\cal H}] \subset {\cal H}, ~~ [{\cal H}, ~ {\cal M}] \subset
{\cal M}, ~~ [{\cal M},~ {\cal M}] \subset {\cal H} .
\label{algebra}
\end{equation}
A Hermitian symmetric space is a symmetric space equipped with a complex
structure. For our purpose, we need only the following properties of
Hermitian symmetric spaces \cite{ford}; for each Hermitian symmetric
space, there exists an element $K$ in the Cartan subalgebra of ${\cal G}$
whose centralizer in ${\cal G}$ is ${\cal H}$, i.e. ${\cal H} = \{ V \in
{\cal G}:~ [V,~ K] = 0 \}$. Also, up to a scaling, $J = \mbox{ad}K = [K, ~ *]$
is a linear map $J: {\cal M} \rightarrow {\cal M}$ satisfying the complex structure
condition $J^{2} = \alpha $ for a constant $\alpha $, or
$[K, ~ [K,~ M]] = \alpha M, ~$ for $ M \in {\cal M}$.
Without loss of generality, we take $\alpha$ to be equal to $-1$.
This complex structure provides a useful identity for $Q$ \cite{park};
\ben
[Q, ~ [Q,~ \partial  Q]] &=& g[K, ~ [K, ~ g^{-1}(\partial  Q)g]]g^{-1}
= g[K, ~ [K, ~ [g^{-1}\partial  g, ~ K]]]g^{-1} \nonumber \\
&=& - g[g^{-1}\partial  g, ~ K]g^{-1} =- \partial  Q ,
\label{comp}
\een
and  along with the symplectic structure, is  the basic ingredient for our 
formulation of the integrable generalized ferromagnet.
In passing, we mention that there exist six types of Hermitian symmetric spaces: 
$SU(p+q)/SU(p)\times SU(q)$, $SO(2n)/U(n)$, $SO(p+2)/SO(p)\times SO(2)$,
$SP(n)/U(n)$ and their noncompact counterparts, and the two exceptional cases.

In application of the above, let us consider the action \cite{park}
\begin{equation}
A= \int dt d^Dx\mbox{ Tr }[2Kg^{-1}\dot g+
\partial_i (gK g^{-1})\partial_i (gK g^{-1})]~~ (i=1,\cdots, D)
\label{actionn}
\end{equation}
where $g$ is a map $g:R^{D+1} \rightarrow G$. The first term
in the action comes from the canonical one form (\ref{theta})
with $x$ replaced by the central element $K$.
This action possesses a local $H$ subgroup 
symmetry so that the physical spin variables take value on the coadjoint 
orbit of the Hermitian symmetric space $G/H$.
The equations of motion can be written in terms of the generalized spin $Q$,
\begin{equation}
\dot Q+\partial_i [Q,~\partial_i Q]=0,
\label{hequation}
\end{equation}
which is the well-known homogeneous LL equation.
In 1+1 dimension, the integrability of the above equation \cite{fadd} arises from the
zero curvature representation;
\be
(\bar{\partial}  - \lambda [Q,~\partial Q] -\lambda^2 Q )
\Psi_{HM} = 0 ,~~~
(\partial   + \lambda  Q)\Psi_{HM} =0,
\label{linear}
\ee
where $\pp = \pp / \pp x , ~ \pb = \pp / \pp t $ and $\lambda $ is an arbitrary
complex constant. These linear equations are overdetermined systems whose
consistency requires the integrability condition;
\ben
0 &=& [\bar{\partial}  - \lambda [Q,~\partial   Q] -\lambda^2 Q , ~
\partial   + \lambda  Q ] \nonumber \\
&=& \lambda (\bar{\partial} Q+\partial [Q,~\partial Q] )
+ \lambda^{2} (\partial Q+[Q, ~ [Q,~ \partial  Q]] ).
\label{zero}
\een
The $\l^{1}$-order term in the second line of the above equation becomes
precisely the LL equation since the $\l^{2}$-order term vanishes
identically due to the complex structure property of the Eq. (\ref{comp}).
Having found the zero curvature representation, one can use the well-
established method \cite{fadd} to calculate the infinitely many
conserved quantities and soliton solutions.

One of the interesting applications of the above representation
is to use  the gauge equivalence of the ferromagnet and the
non-linear Schr\"odinger (NS) model \cite{zakh} to derive 
the generalized NS equation from the generalized LL equation.
The explicit procedure and the subsequent equation can be found
in Ref. \cite{park}.
In the same Reference, such a gauge equivalence was also used  to  relate the 
conserved integrals of both models and explicit conserved quantities in both models
were found.

Next, we show that the reduced action of the Eq.(\ref{actionn}) 
on the coadjoint orbit  
describes the generalized Hamiltonian dynamics \cite{dira}.
In order to be explicit,  we restrict to the 
$CP(N-1) = SU(N)/(SU(N-1) \times U(1))$ case where the element $K$ 
in the Cartan subalgebra is given by 
$K=(i/N) \mbox{diag}(N-1,-1,\cdots, -1)$. Now introduce a parameterization of the 
group element $g$ of $SU(N)$ by an $N$-tuple, $ g = (Z_1, Z_2,\cdots, Z_N);
\ Z_p\in {\bf C}^N$ $(p=1,\cdots,N)$, such that
\begin{equation}
\bar Z_pZ_q=\delta_{pq}, \quad \mbox{det}
(Z_1, Z_2,\cdots, Z_N) =1.
\label{cond}
\end{equation}
Then the generalized spin $Q$ is given by
\begin{equation}
Q=iZ_1\bar Z_1-iI .
\label{isosp}
\end{equation}
All other $Z_p$'s with $p=2,\cdots,N$ disappear in the expression 
of $Q$ due to the particular form of $K$.
In terms of the Fubini-Study coordinate 
$\psi_\alpha (\alpha=1,2,\cdots, N-1$) \cite{alek};
\begin{equation}
z_\alpha=\frac{\psi_\alpha}{\sqrt{1+\vert\psi\vert^2}},\quad
z_0=\frac{1}{\sqrt{1+\vert\psi\vert^2}}~;~
Z_1^T=(z_0,z_1,\cdots,z_{N-1})
\label{slp}
\end{equation}
we have an equivalent expression of $Q$ in component, 
\begin{equation}
Q^A(\psi,\bar\psi)=-2i\sum_{p,q=0}^{N-1}\bar z_p(T^A)_{pq}z_q .
\label{spinfunction}
\end{equation}
Substituting the above expression into the action (\ref{actionn}),
we obtain a reduced action on the $CP(N-1)$ orbit (up to a total
derivative term and trivial rescaling),
\begin{equation}
A=\int dt dx\left(2i\frac{\bar \psi_\alpha\dot\psi_\beta}
{1+\vert\psi\vert^2}- g_{\alpha\beta}\partial \psi_\alpha
\partial \bar\psi_\beta\right),
\end{equation}
where $g_{\alpha\beta}$ is the Fubini-Study metric on $CP(N-1)$,
\begin{equation}
g_{\alpha\beta}=\frac{(1+\vert\psi\vert^2)\delta_{\alpha\beta}
-\bar\psi_\alpha\psi_\beta}{(1+\vert\psi\vert^2)^2}.
\end{equation}
Note that the first term in the above action can be written as
$\int dx \theta$ where $\theta$ is the canonical one-form
on $CP(N-1),~ \theta=2i\partial_{\psi} \log(1+\vert\psi\vert^2)d\psi $
and the classical dynamics can be described
by a generalized Hamiltonian dynamics \cite{dira}
with the Hamiltonian given by
\begin{equation}
H=\int dx g_{\alpha\beta}\partial \psi_\alpha
\partial \bar\psi_\beta.
\end{equation}
The Poisson bracket  defined by the
inverse matrix $\omega^{\alpha\beta}=-ig^{\alpha\beta}$
of the symplectic two-form $\omega=d\theta$,
\begin{equation}
\{F(\bar\psi,\psi),G(\bar\psi,\psi)\}=-\frac{i}{2}
\int dx ~g^{\alpha\beta}\left(\frac{\delta F
}{\delta \bar \psi_\alpha(x)}\frac{\delta G}{\delta\psi_\beta(x)}-
\frac{\delta G}{\delta \bar\psi_\alpha(x)}
\frac{\delta F}{\delta \psi_\beta(x)}\right)
\label{pbracket}
\end{equation}
with the inverse Fubini-Study metric $g^{\alpha\beta}=(1+\vert\psi\vert^2)
(\delta_{\alpha\beta}+\bar\psi_\alpha\psi_\beta)$ reproduces the generalized spin algebra, 
the Eq. (\ref{spin}) as expected.
Also the Hamiltonian equation of motion gives the generalized 
LL equation (\ref{hequation});
\begin{equation}
\dot Q^A=\{H, Q^A\}=-f^{ABC}Q^B\partial ^2Q^C.
\end{equation}

Let us  consider the above model (\ref{actionn}) in the (2+1)-dimensional case 
and couple with the Chern-Simons gauge fields
to study the self-dual Chern-Simons solitons \cite{jackiw}.
We introduce gauge fields $A_{\m}$ which gauges the left 
multiplication of group $G ~ ; ~g \rightarrow g^{'}g$; 
\begin{equation}
S= \int dt d^2x\left\{[\mbox{ Tr }(2Kg^{-1}D_t g+
D_i(gK g^{-1})D_i (gK g^{-1}))]-V(gKg^{-1})+
{\cal L_{CS}}\right\}.
\label{action}
\end{equation}
The covariant derivative is defined on fundamental and adjoint 
representations by
\begin{equation}
D_\mu g =\partial_\mu g +A_\mu g,~~~A_\mu=A_\mu^A T^A , ~~~ 
D_{\m }(gKg^{-1} ) = D_{\m }Q = [ \pp_{\m } + A_{\m } ~ , ~ Q] .
\end{equation}
The potential $V(gKg^{-1})$ is given by
\begin{equation}
V=\frac{1}{2}I^{AB}Q^AQ^B
\label{potential}
\end{equation}
where $I^{AB}$ is a constant symmetric matrix measuring the anisotropy of 
the system \cite{kose}. We assume that the dynamics of gauge fields is 
governed by the Chern-Simons action ${\cal L}_{CS}$:
\begin{equation}
{\cal L}_{CS}=-\kappa\epsilon^{\mu\nu\rho}\mbox{Tr}
(\partial_\mu A_\nu A_\rho+\frac{2}{3}A_\mu A_\nu A_\rho).
\end{equation}
Then, the equations of motion in terms of the generalized spin $Q$ are the 
gauged planar LL equation;
\begin{equation}
D_{t} Q+D_i [Q,~D_i Q]+[\bar Q,Q]=0
\label{gequation}
\end{equation}
with $\bar Q=I^{AB}Q^AT^B$. The Gauss's law constraint is given by
\begin{equation}
G^A=  \frac{\kappa}{2}\epsilon^{ij}F_{ij}^A-Q^A =0.
\end{equation}
Note that the above constraint is of a similar type with the one 
which appears in the well-known  nonrelativistic Chern-Simons gauged 
NS model \cite{jak1}.

The Hamiltonian is given by
\be
H=\int d^2x {\cal H}=
\int d^{2}x [ \frac{1}{2}(D_{i}Q^A)^2 + V(Q^A) ] .
\ee
The useful identity  Eq. (\ref{comp}), which still holds
with $\partial Q$ being replaced by $D_iQ$, brings the 
Hamiltonian $H$ into the Bogomol'nyi type;
\be
H =
\int d^{2}x [\frac{1}{4}( D_{i}Q^A\pm\epsilon_{ij}
[Q,  D_{j}Q ]^A)^2 + V(Q^A) ]
\pm \frac{1}{2}\epsilon_{ij}F_{ij}^AQ^A ]
\pm 4\pi T ,
\label{bogh}
\ee
where the topological charge $T$ is defined by
\begin{equation}
T =\frac{1}{8\pi}
\int d^2x [\epsilon_{ij}Q^A 
[\partial_iQ,\partial_jQ]^A-2\epsilon_{ij}\partial_i (Q^AA_j^A)].
\label{tcharge}
\end{equation}
Thus, the energy is bounded below by the topological charge $T$ when the 
potential $V$ is chosen such that
\be
V \pm \frac{1}{2}\epsilon_{ij}F_{ij}^AQ^A = 0 .
\ee
Or, upon imposing the Gauss's law constraint, it is equivalent to 
choosing the constant matrix 
\begin{equation}
I^{AB}=\mp\frac{2}{\kappa}\delta_{AB}.
\label{isotropic}
\end{equation}
The minimum energy arises when the spin variable satisfies the 
first order self-duality equation,
\begin{equation}
D_iQ=\mp \epsilon_{ij}[Q, D_jQ].
\label{selfdual}
\end{equation}
Note that with the choice (\ref{isotropic}), the potential 
(\ref{potential}) reduces to a constant. Also, in the absence of gauge fields 
the self-duality equation (\ref{selfdual}) is precisely 
that of two dimensional instantons in the principle chiral model which has been classified 
according to each symmetric spaces \cite{pere}. 

Vortices can arise in our model, if we take the gauge group  
to be the maximal torus subgroup of $H$ and introduce gauge invariant 
terms to the action which induce vacuum symmetry breaking. Explicitly, we
take $H^{a} (a = 1, \cdots , \mbox{rank} (H))$ to be generators of the maximal 
torus group and add to the action Eq. (\ref{action}) a uniform
background "charge" term
\begin{equation}
\D S=\int dtd^{2}x A_o^av^a,
\end{equation}
where each $v^a$ is a constant and the sum is taken over $a=1, \cdots, 
\mbox{rank} (H)$. 
Then, the gauge fields $A_{\m } = A_{\m}^{a}H^{a}$ and the Chern-Simons 
action reduces to a sum of Abelian Chern-Simons terms,
\begin{equation}
{\cal L}_{CS}=\frac{\kappa}{2}\epsilon^{\mu\nu\rho}
\partial_\mu A^a_\nu A^a_\rho ~ .
\end{equation}
The Gauss's law is replaced by
\begin{equation}
\frac{\kappa}{2}\epsilon_{ij}F^a_{ij} =Q^a-v^a.
\label{agauss}
\end{equation}
Also, we have the topological charge replacing Eq. (\ref{tcharge}) 
\begin{equation}
T=\frac{1}{8\pi}\int d^2x[\epsilon_{ij}Q^A [\partial_iQ, \partial_jQ]^A
+2\epsilon_{ij}\partial_i((v^a-Q^a)A^a_j)] .
\end{equation}
Assuming the potential $V$ to be of the form
\begin{equation}
V(gKg^{-1})=\frac{1}{2}\sum_a I^a(Q^a-v^a)^2 ,
\label{pot2}
\end{equation}
we find that the Bogomol'nyi bound is established with the choice
\begin{equation}
I^1=\cdots =I^{N-1}=\mp\frac{2}{\kappa}.
\end{equation}
Note that the potential Eq. (\ref{pot2}) is nontrivial unlike the 
previous case and the nonvanishing constants $v^{a}$ breaks 
the symmetry of the vacuum spontaneously.

Let us consider an explicit example with $CP(N-1) = 
SU(N)/(SU(N-1) \times U(1))$.
We choose the standard expression for $T^A$'s:
$T^A=i\lambda^A/2$ where $\lambda^A$ is the  $SU(N)$ Gell-Mann
matrices. The Cartan subalgebra generators $H^{a}$ 
generating the maximal torus group of $SU(N-1) \times U(1)$ 
are given by $N-1$ diagonal matrices 
\begin{equation}
H^a_{pq}=i(\sum_{k=1}^{a}\delta_{ik}\delta_{jk}
-a\delta_{i,a+1}\delta_{j,a+1})/\sqrt{2a(a+1)}~;~ a = 1, \cdots, N-1 .
\end{equation}
Using the complex notation;
 $z = x+iy, \bar z = x-iy$,
$A_z  = \frac{1}{2}(A_1 - iA_2), A_{\bar z} =
\frac{1}{2} (A_1 + iA_2)$,  and
$D_z = \frac{1}{2}(D_1-iD_2),
D_{\bar z} = \frac{1}{2}(D_1+iD_2)$, we obtain
an alternative expression of the self-duality equation, 
\begin{equation}
D_zQ=\mp i[Q, D_zQ].
\label{selfdualeq}
\end{equation}
With the parameterization of $Q$ as in Eq. (\ref{spinfunction}), the self-duality 
equation (\ref{selfdualeq}) for the plus sign case becomes 
a set of $N-1$ equations:
In terms of a notation
\begin{equation}
D_-^{\alpha } \equiv \partial_z+\frac{i}{2}(A_z^1+\frac{1}{\sqrt{3}}A_z^2
+\cdots +\frac{1}{\sqrt{(\alpha-1)(\alpha-2)/2}}A^{\alpha-2}_z+
\frac{\alpha}{\sqrt{\alpha(\alpha-1)/2}}A^{\alpha-1}_z),
\end{equation}
we have
\begin{equation}
D_-^\alpha\bar\psi_\alpha=0 ~ ; ~ \a = 1, \cdots , N-1.
\label{equa1}
\end{equation}
Similarly, for the minus sign case, we have
\begin{equation}
D_+^\alpha \equiv \partial_z-\frac{i}{2}(A_z^1+\frac{1}{\sqrt{3}}A_z^2
+\cdots +\frac{1}{\sqrt{(\alpha-1)(\alpha-2)/2}}A^{\alpha-2}_z+
\frac{\alpha}{\sqrt{\alpha(\alpha-1)/2}}A^{\alpha-1}_z),
\end{equation}
and
\begin{equation}
D_+^\alpha\psi_\alpha=0 ~ ; ~ \a = 1, \cdots , N-1.
\label{equa2}
\end{equation}
The Gauss's law constraint Eq. (\ref{agauss}) is given by
\begin{equation}
\partial_z A^a_{\bar z}-\partial_{\bar z}A^a_z=
Q^a(\psi,\bar\psi)-v^a.
\label{agausss}
\end{equation}

In the $CP(1)$ case, we have only one complex $\psi $ which we parameterize by
\be
\bar\psi=\rho\exp(i\phi)
\ee
where $\rho $ is real and the phase $\phi $ is a real multi-valued function.
Then, Eq. (\ref{equa1}) can be solved for the gauge field $A$  and
Eq. (\ref{agausss}) reduces to a vortex-type equation;
\begin{eqnarray}
A_i&=& \epsilon_{ij}\partial_j\log\rho-\partial_i\phi
\nonumber\\
\nabla^2 \log\rho&+&\epsilon_{ij}\partial_i\partial_j\phi
= \frac{1}{\kappa}(v-\frac{1-\rho^2}{1+\rho^2}).
\end{eqnarray}
The derivative term $ \epsilon_{ij}\partial_i\partial_j\phi $ is identically 
zero except at the zeros of $\bar\psi $ where the multi-valuedness of $\phi $ 
results in the Dirac delta function(see, for example, \cite{taubes}).
A numerical analysis suggests that these vortex-type equations  
possess vortex solutions which exhibit anyonic property 
and show a rich structure depending on the value of $v$ \cite{oh}. 
Higher $N$ case was also treated in the Ref. \cite{park1}.
                        
In conclusion, we have shown that  each Hermitian symmetric space
plays an essential role in the formulation of integrable generalized Heisenberg
ferromagnet in 1+1 dimension, and for the self-dual Chern-Simons solitons
and vortices. It would be  interesting  to extend the above idea to the 
relativistic field theory  and also to investigate the quantization problem.

\begin{center}
{\bf ACKNOWLEDGEMENTS}
\end{center}
I would like to thank Y. Kim, Q-H. Park and C. Rim for useful
discussions and C. Lee for encouragement. This work is supported by the KOSEF
through the CTP at SNU and the project number
96-0702-04-01-3.


\begin{thebibliography}{99}
\bibitem{park} P. Oh and Q-H. Park, Phys. Lett. B. {\bf 383}
 (1996) 333.
\bibitem{park1}P. Oh and Q-H. Park, hep-th/9612063.
\bibitem{kiri} A. A. Kirillov, {\it Elements of the Theory of
 Representations} (Springer-Verlag, 1976).
\bibitem{ford} A. P. Fordy and P. P. Kulish, Commum. Math. Phys. {\bf 89}
(1983) 427; S. Helgason,  {\it Differential geometry, Lie groups and 
Symmetric Spaces} 2nd ed. (New York, Academic Press, 1978).
\bibitem{arno} R. Abraham and J. E. Marsden,  
{\it Foundations of Mechanics} (Addison Wesley, New York, 1978).
\bibitem{ohnucl} P. Oh, Nucl. Phys. B {\bf 462} (1996) 551.
\bibitem{bak} See also D. Bak, R. Jackiw and S.-Y. Pi, Phys. Rev. D 
{\bf 49} (1994) 6778 in which the relation is proved by an explicit
use of the projection operator.  
\bibitem{fadd} L. D. Faddeev and L. A. Takhtajan,
{\it Hamiltonian Methods in the Theory of Solitons} (Springer-Verlag,
Berlin, 1987) and references therein.
\bibitem{zakh} V. E. Zakharov and L. A. Takhtadzhyan, Theor. Math. Phys.
{\bf 38} (1979) 17.
\bibitem{alek} T. Lee and P. Oh, Phys. Lett. B  {\bf 319}  (1993) 497.
\bibitem{dira} P. A. M. Dirac,  {\it Lectures on Quantum Mechanics}
(Yeshiva Univ., New York, 1964);
 L. D. Faddeev and R. Jackiw, Phys. Rev. Lett.  {\bf 60} (1988) 1692.
\bibitem{jackiw} G. Dunne,  {\it Self-Dual Chern-Simons Theories} (Springer,
 Berlin, 1995)  and  references therein.
\bibitem{kose} A. M. Kosevich, B. A. Ivanov and A. S. Kovalev, Phys. Rep.
 {\bf 194} (1990) 117.
\bibitem{jak1} R. Jackiw and S.-Y. Pi, Phys. Rev. Lett. {\bf 64} (1990) 2969;
Phys. Rev. D {\bf 42} (1990) 3500; {\bf 48}  (1993) 3929(E).
\bibitem{pere} A. M. Perelomov, Phys. Rep. {\bf 146} (1987) 135. 
 \bibitem{taubes} A. Jaffe and C. Taubes, {\it Vortices and Monopoles} 
(Birkh\"{a}user, Boston, 1980).
\bibitem{oh} Y. Kim and P. Oh, to be published.
\end{thebibliography}
\end{document}